\documentclass[twocolumn,showpacs,preprintnumbers,amsmath,amssymb,floatfix,nofootinbib]{revtex4-1}
\usepackage{color}   
\usepackage{graphicx}
\usepackage{dcolumn}
\usepackage{bm}
\usepackage{slashed}
\usepackage{float}

\begin{document}

\def\bbox #1{\hbox{\boldmath${#1}$}}
\def\bb    #1{\hbox{\boldmath${#1}$}}
 \def\oo    #1{{#1}_0 \!\!\!\!\!{}^{{}^{\circ}}~}  
 \def\op    #1{{#1}_0 \!\!\!\!\!{}^{{}^{{}^{\circ}}}~}
\def\blambda{{\hbox{\boldmath $\lambda$}}}
\def\eeta{{\hbox{\boldmath $\eta$}}}
\def\bxi{{\hbox{\boldmath $\xi$}}}

\title{ Shells in a Toroidal Nucleus in the Intermediate Mass Region }

\author{Cheuk-Yin Wong}
\email{wongc@ornl.gov}
\affiliation{Physics Division, 
Oak Ridge National Laboratory, 
Oak Ridge, TN 37831, USA}

\author{Andrzej Staszczak}
\email{andrzej.staszczak@poczta.umcs.lublin.pl}
\affiliation{Institute of Physics, Maria Curie-Sk{\l}odowska University,
pl.~M.~Curie-Sk{\l}odowskiej 1, 20-031 Lublin, Poland}

\date{\today}

\begin{abstract}

 Attention is fixed on shells in toroidal nuclei in  the intermediate mass
 region using a toroidal single-particle potential.  We find that
 there are toroidal shells in the intermediate mass region with large
 single-particle energy gaps at various nucleon numbers located at
 different toroidal deformations characterized by the aspect ratios of
 toroidal major to minor radius.  These toroidal shells provide extra
 stability at various toroidal deformations.  Relative to a toroidal
 core, Bohr-Mottelson spin-aligning particle-hole excitations may be
 constructed to occupy the lowest single-particle Routhian energies to
 lead to toroidal high-spin isomers with different spins. Furthermore,
 because a nucleon in a toroidal nucleus possesses a vorticity quantum
 number, toroidal vortex nuclei may be constructed by making
 particle-hole excitations in which nucleons of one type of vorticity
 are promoted to populate un-occupied single-particle orbitals of the
 opposite vorticity.  Methods for producing toroidal high-spin isomers
 and toroidal vortex nuclei are discussed.

\end{abstract}

\pacs{   21.10.Pc,  21.60.Cs  }

\maketitle
\section{Introduction}
\label{intro}
Wheeler suggested that under appropriate conditions, the nuclear fluid
may assume the shape of a torus \cite{Whe50}.  Favorable conditions
include nuclear shell effects \cite{Won72,Won73}, large Coulomb
energies \cite{Won72,Won73,Won78}, large nuclear angular momenta
\cite{Won78a,Won85,Zha86,Wil86}, nuclear collision dynamics
\cite{Xu93,Xu94}, and combinations thereof.  Although the basic ideas
on the conditions favorable for toroidal configurations were presented
many decades ago
\cite{Won72,Won73,Won78,Won78a,Won85,Zha86,Wil86,Xu93,Xu94}, the subject
matter gains renewed interest recently because powerful theoretical and
experimental tools are now readily available
{\cite{War07,Vin08,Sta09,Zha10,Ich12,Ich14,Sta14,Sta15,Sta15a,Sta16,Men13,Ren17,
    Zhao10,Lal05,Kos17,Sta17,Kos18,Cao18,Afa18,Hag94,Lar96,Wue09,Wad04,Lin14,
    Cas87,Ono99,Cha10,Man10,Lac04}.  The interest is heightened by
  recent experimental evidence for the presence of resonances at high
  excitation energies in the 7$\alpha$ disassembly of $^{28}$Si which
  may suggest the production of toroidal high-spin isomers predicted 
in many theoretical calculations  \cite{Cao18}.  Should these experimental results be confirmed by
  further studies, toroidal nuclei would potentially be interesting
  objects of study because of their new form of geometry, new toroidal
  shells and magic numbers, new types of yrast high-spin states, new
  toroidal nuclei species in different mass regions, new probes of
  nuclear energy density functional and nuclear equation of state in a
  new density regime, and new possible doorways to energy-producing
  mechanisms.

It is instructive to see how nuclear shell effects and the alignment
of single-particle spins along the symmetry axis can constrain the
nucleus to assume a toroidal shape.  We characterize the toroidal
deformation of a toroidal nucleus by the aspect ratio $R/d$ of the
major radius $R$ to the minor radius $d$.  For light nuclei, there are
large energy gaps in toroidal single-particle energies in an extended region
of toroidal deformations.  These energy gaps give rise to ``toroidal
shells'' at ``magic" nucleon numbers $N$=$2(2m+1)$, with integers
$m\ge 1$ \cite{Won72,Won73}.  The extra stability associated with
toroidal shells \cite{Bra73} leads to toroidal local energy minima for
many light nuclei.  Toroidal excited (diabatic) states have been
predicted for Ca to Ge, with mass numbers 40$\lesssim$$ A$$\lesssim
70$ using the Strutinsky shell correction method \cite{Won72,Won73},
and for $^{24}$Mg \cite{Zha10} and $^{28}$Si \cite{Cao18} using a
self-consistent relativistic mean-field theory.  Relative to a
toroidal core, spin-aligning Bohr-Mottelson particle-hole excitations
occupying the lowest Routhian single-particle energies ~\cite{Boh81}
can be constructed to yield a toroidal nucleus with a spin as an yrast
state, by promoting nucleons with angular momentum aligned opposite to
a chosen symmetry $z$-axis to populate orbitals with angular momentum
aligned along the symmetry $z$-axis
~\cite{Ich12,Ich14,Sta14,Sta15,Sta15a,Sta16}.  A spinning toroidal nucleus
possesses an effective ``rotational'' energy which tends to expand the
toroid, whereas the energy associated with the nuclear bulk properties
tends to contract the toroid.  The balance between the two energies
gives rise to a local toroidal energy minimum \cite{Won78}.
Self-consistent calculations have been carried out to locate toroidal
high-spin isomers as yrast states in the light mass
region~\cite{Ich12,Ich14,Sta14,Sta15,Sta15a,Sta16,Cao18}.

For the heavy and superheavy nuclei there is similarly a toroidal
shell region with negative shell corrections \cite{Won72,Won73}.  The
toroidal configuration is further favored by large Coulomb energies.
As a consequence, many toroidal nuclei have been theoretically located
in the superheavy region
\cite{War07,Vin08,Sta09,Kos17,Sta17,Kos18,Afa18}.  In situations where
the $I$=0 state in the toroidal nucleus $^{304}$120 are not stable
adiabatically, the spin-aligning Bohr-Mottelson particle-hole
excitations generate yrast high spin states to lead to adiabatic local
energy minima at various spin values \cite{Kos17,Sta17}.  The
neighboring even-even $N$=184 isotone nuclei with $Z$=118 and $122$,
as well as the $Z$=120 isotopes with $N$=182 and 186, also possess
toroidal high spin isomers at various spins and toroidal deformations
\cite{Kos18}.

Along with theoretical predictions
\cite{Won72,Won73,Zha10,Ich12,Ich14,Sta14,Sta15,Sta15a,Sta16,War07,Vin08,Sta09,Kos17,Sta17,Kos18,Afa18}
and experimental investigations \cite{Cao18}  in
the light and heavy mass regions, it is natural to ask whether
toroidal nuclei may also be possible in the intermediate mass region.
In this regard, our knowledge of toroidal nuclei in this mass 
region is scanty. 
We wish to examine here toroidal shells in the intermediate mass region
with 30$\lesssim$$N$(or $Z$)$\lesssim$96 where negative shell
corrections are present over an extended region of nucleon numbers and
toroidal deformations \cite{Won73}.  A large number of toroidal shells
at various toroidal deformations are expected.  We would like to know 
 whether there are some regularities in
the shell structure,
how frequent do the toroidal shells occur, what are their nucleon numbers, and where their
corresponding toroidal deformations are located.
  Furthermore,
Bohr-Mottelson spin-aligning particle hole excitations relative to a
toroidal core can lead to high-spin toroidal isomers and these
high-spin isomers as yrast states have longer lifetimes and a better
chance of being detected \cite{Cao18}.  We would like to examine the
possibilities of toroidal high-spin isomers in this mass region by
studying the shell structure of single-particle Routhian energies as a
function of the cranking frequency.  Such information will guide our
intuition and help our search for toroidal excited states and
toroidal high-spin isomers in future experiments and mean-field calculations in the
intermediate mass region.

In the present survey over a large extended multi-dimensional space of
nucleon numbers $N$, toroidal deformations $R/d$, and nuclear spins
$I$=$I_z$, it is convenient to use a simple harmonic oscillator
toroidal shell model for which the relevant results can be readily
obtained.  Furthermore, the simple toroidal shell model has the
desirable property that many physical properties can be clearly
exhibited.  In particular, from the single-particle wave functions for
the two-dimensional harmonic oscillators in the meridian plane, one
finds that a nucleon in a toroidal nucleus possesses a vorticity
quantum number $\Lambda_\perp$  associated with a circulating current around the
poloidal angle $\theta$ ( Fig.\ 1).  We propose the possibility of toroidal
vortex nuclei by making vortex-creating single-particle particle-hole
excitations that promote nucleons of one type of vorticity from
occupied orbitals to populate un-occupied single-particle orbitals of
the opposite vorticity.  The resultant nucleus will have a net
non-zero vorticity $\Lambda_\perp$ as shown in Fig \ref{fig1}.  Because
vorticity is quantized and cannot be easily destroyed, the presence of
a net vorticity may enhance the stability of the toroidal vortex
nucleus.

\begin{figure} [t]
\centering
\includegraphics[scale=0.45]{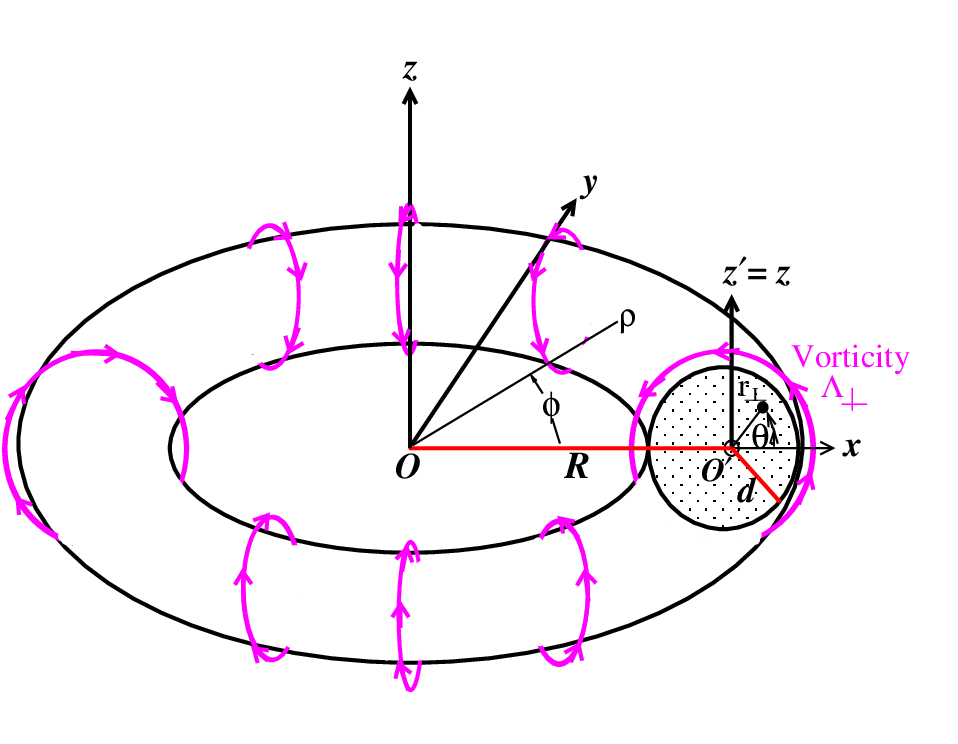} 
\caption{(Color online) The cylindrical coordinates $(\rho, z, \phi)$ and toroidal
  coordinates  $(r_\perp,\theta,\phi)$
    used for the description
  of a toroidal nucleus with a major radius $R$, a minor radius $d$,
  and a vorticity $\Lambda_\perp$.  Here $\phi$ is the toroidal angle,
  $\theta$ is the poloidal angle, and $r_\perp=\sqrt{(\rho-R)^2+z^2}$.
  Some of the poloidal vortex flow lines are schematically exhibited.
}
\label{fig1}
\end{figure}

It should however be noted that the vorticity quantum number is
non-zero only if the principal quantum number $n$ of the
two-dimensional harmonic oscillator wave function for the nucleon
motion on the meridian plane is greater than or equal to unity.  This
excludes the light toroidal nuclei from the possibility of becoming
toroidal vortex nuclei.

Experiments have been performed recently to search for a toroidal
high-spin isomer in $^{28}$Si with spin $I$=$I_z$=44 predicted earlier
in \cite{Sta14} by colliding $^{28}$Si at 35 MeV/A on a fixed $^{12}$C
target \cite{Cao18}.  At the predicted energy region,
 a total of
three sharp resonances 
instead of just a single resonance
have been
observed and interpreted as possible toroidal high-spin isomers with
spins $I$=$I_z$=28, 36, and 44 in the excited $^{28}$Si system.  Subsequent 
relativistic  mean-field calculations  provide additional theoretical support 
for the presence of these states \cite{Men13,Ren17,Zhao10}.  We would like to understand
in a simple mechanical way how high-spin nuclei can be produced in the
binary products of such reactions.  We would also like to investigate
the process of punching of a small nucleus through a larger nucleus
\cite{Won73} for the generation of vorticities, if only heuristically.

This paper is organized as follows.  In Section II, we introduce the
toroidal single-particle potential and evaluate the single-particle
eigenenergies analytically as shown in the Appendix.  In Section III,
we display the single-particle state energies in the intermediate mass
region, and infer the toroidal shell magic number $N$ in a region of
low single-particle energy densities as a function of the toroidal
deformation $R/d$.  In Section IV, we examine how one can carry out
the spin-aligning Bohr-Mottelson particle-hole excitations to lead to
toroidal high-spin isomers as yrast states in the intermediate mass
region.  The set of nucleon numbers and spin quantum numbers
 for the favorable configurations at $R/d$=2.9  are presented
as an example.  In Section V, we make the coordinate transformation from
the Cartesian-like coordinates of $(\rho$$-$$R,z)$ to the polar
coordinates $(r_\perp$,$\theta)$ on the meridian plane in order to
exhibit explicitly the vorticity quantum number $\Lambda_\perp$.  In
Section VI, we show how we can construct a vortex nucleus by
vortex-creating single-particle particle-hole excitations.  In Section
VII, we examine some experimental methods in the production of
toroidal high-spin isomers and toroidal vortex nuclei.  Section VIII
concludes the present discussions.

\section{Toroidal single-particle potential}

The study of the shell structure of a toroidal nucleus necessitates
the use of a toroidal single-particle potential, which has a shape
similar to the toroidal density distribution.  Accordingly, we assume
a single-particle toroidal potential of the form \cite{Won73}
\begin{eqnarray}
V_0(\rho,z) = \frac{1}{2} m \omega_\perp^2 (\rho -R)^2 +\frac{1}{2} m \omega_\perp^2  z^2,
\end{eqnarray}
where the $z$-axis is the symmetry axis, $\rho=\sqrt{x^2+y^2}$ as
shown in Fig.\ \ref{fig1}.  The quantity $m$ is the nucleon rest mass
and $\omega_\perp$ is the harmonic oscillator frequency related to the
aspect ratio $R/d$ by \cite{Won73}
\begin{eqnarray}
\hbar \omega_\perp&=& \left (   \frac{3\pi R/d}{2}\right ) ^{1/3}\hbar \oo \omega,
\end{eqnarray}
where
\begin{eqnarray}
\hbar \oo \omega& =& 41 {\rm MeV}/ A^{1/3}.
\label{eq3}
\end{eqnarray}
The above equation for $\hbar \oo \omega$ needs to be modified.  In
earlier HFB calculations \cite{Sta14}, the average toroidal
nuclear matter density $\rho_{\rm toroidal}$ is approximately 2/3 of the
nuclear matter density $\rho_{0}$ of the spherical nucleus with the
same mass number.  Because the mean-field potential is proportional
approximately to the nuclear matter density, we need to include an
additional factor $\rho _{\rm toroidal}/\rho_0$ in Eq.\ (\ref{eq3}) to
give
\begin{eqnarray}
\hbar\oo  \omega =( 41 {\rm MeV}/ A^{1/3})(\rho _{\rm toroidal}/\rho_0).
\label{4}
\end{eqnarray}
We have $(\rho _{\rm toroidal}/\rho_0)$$\sim$$ 0.64$ for $A$$\sim$$
40$ \cite{Sta14}, and $(\rho _{\rm toroidal}/\rho_0)$$\sim$$ 1.0$ for
superheavy nuclei \cite{Sta17}, we can therefore take the average
value of $(\rho _{\rm toroidal}/\rho_0)$$\sim$0.82 for our
intermediate mass region of $A$$\sim$60-160, where we shall use
$A$=110 in Eq.\ (\ref{4}) for numerical purposes.

With the inclusion of  the spin-orbit interaction,  the single-particle potential is
\begin{eqnarray}
H= -\frac{\hbar ^2}{2m} \nabla^2 + V_0(\rho,z) - \frac{2\kappa \hbar}{m \oo \omega} \bb s \cdot \left ( \nabla V_0 \times \bb p \right ) ,
\label{eq5}
\end{eqnarray}
where $\kappa$ is a dimensionless parameter for which Nilsson gave the
value of $\kappa=0.06$ \cite{Nil69}.  We choose the spin $\bb s$ to be
diagonal along the symmetry $z$-axis and neglect the small
contribution from the off-diagonal spin-orbit interaction.  We carry
out our analytical calculations in the large major radius
approximation in which $R\gg d$, expand $\rho$ about $R$ in power of
$q$ = $(\rho-R)$, and keep terms up to the second order in $q/R$.  The harmonic
oscillator potential can be solved analytically as shown in the
Appendix.  We get the single-particle energies for the single-particle
state $|n_\rho n_z \Lambda_z \Omega_z\rangle$
\begin{eqnarray}
 \epsilon(n_\rho n_z \Lambda_z \Omega_z)&& =(n_\rho+\frac{1}{2})\hbar \omega_\perp' + (n_z+\frac{1}{2})\hbar \omega_\perp
  \nonumber\\
&&+  \frac{\hbar ^2}{2m} \frac{\Lambda_z^2-\frac{1}{4}}{R^2} +a_0,
\label{eq6}
\end{eqnarray}
where $n_\rho$ and $n_z$ are the quantum numbers for harmonic
oscillations in the $\rho$ and $z$ directions respectively,
$\Lambda_z$ is the azimuthal angular momentum quantum number,
$\Omega_z=\Lambda_z+s_z=\pm |\Omega_z|$,
\begin{eqnarray}
\omega_\perp'^2&&=\omega_\perp^2(1+a_2), 
\label{21}
\end{eqnarray}
\begin{eqnarray}
a_2&=& \frac{1}{m\omega_\perp^2} \left \{  \frac{\hbar ^2}{2m} \frac{\Lambda_z^2-\frac{1}{4}}{R^2}\frac{6}{ R^2} + \frac{2\kappa (\hbar \omega_\perp)^2}{\hbar \oo  \omega} 
s_z L_z\frac{2}{R^2}\right \}\!,~~
\label{22}
\\
a_0&=& \frac{1}{2} m \omega_\perp ^2 q_0^2 
+ \frac{\hbar ^2}{2m} \frac{\Lambda_z^2-\frac{1}{4}}{R^2}
\left (  -\frac{2q_0}{R}
+\frac{3q _0^2}{ R^2}\right )
\nonumber\\
& &~~~~~~~~~~ -\frac{2\kappa (\hbar \omega_\perp)^2}{\hbar \oo  \omega} 
s_z L_z
\left ( \frac{q_0}{R} - \frac{q_0^2}{R^2} \right ),
\label{23}
\\
q_0 \!&=& \! \frac{1}{m\omega_\perp^2(1\!+\!a_2)} \! \left \{ \!
\frac{\hbar ^2}{2m} \frac{\Lambda_z^2\!-\!\frac{1}{4}}{R^2}\frac{2}{ R} \!+\! \frac{2\kappa (\hbar \omega_\perp)^2}{\hbar \oo  \omega} 
\frac{s_z L_z}{R}\!\! \right \}\!.~~
\label{24}
\end{eqnarray}

\section{Toroidal single-particle states }

The single-particle state energies in Eq.\ (\ref{eq6}) lead to the
level diagram as a function of the toroidal deformation $R/d$ shown in
Fig.\ \ref{fig2}. A toroidal shell $(N,$$R/d)$ is characterized by the nucleon
number $N$, for which the single-particle states have an energy gap at
the toroidal deformation $R/d$.  We indicate the location of a
toroidal shell $(N,R/d)$ by a bracketed number $(N)$ at its
corresponding toroidal deformation $R/d$ in Fig.\ \ref{fig2}.

We find that toroidal shells are numerous in number in light and
intermediate mass nuclei.  As the nucleon number increases beyond the
light mass region, nucleons populate states with
$n$=$n_\rho$+$n_z$$\ge$1 and the density of single-particle states
initially becomes very dense.  The density however becomes sparse as
the nucleon number and toroidal deformation increase.  As indicated in
Fig.\ \ref{fig2}, large energy gaps of single-particle energy levels occur to
give rise to toroidal shells for many nucleon numbers at various
toroidal deformations.  The toroidal shells  $(N,R/d)$  in the intermediate mass region with
30$\lesssim$$N$(or~$Z$)$\lesssim$96
are listed in Table \ref{tab1}.  

We note that the toroidal shells  occur in a rather regular pattern as a result of the interplay 
among three different energies scales as is evident from Eq.\ (\ref{eq6}).  There is first  of all  the gross structure energy scale of 
$\hbar \omega_\perp$ of order 10 MeV
 that increases with
$R/d$.  There is the much smaller energy scale of
$\hbar^2/2mR^2$ of order 1 MeV with   single-particle orbiting
energies   $\Lambda_z^2 \hbar^2/2mR^2$.
And finally, there is the small finer-structure arising from the spin-orbit interactions.
As a result of the interplay between these three energy scales, we note the 
following regular structure:

\begin{figure} [t]
\centering
\includegraphics[scale=0.65]{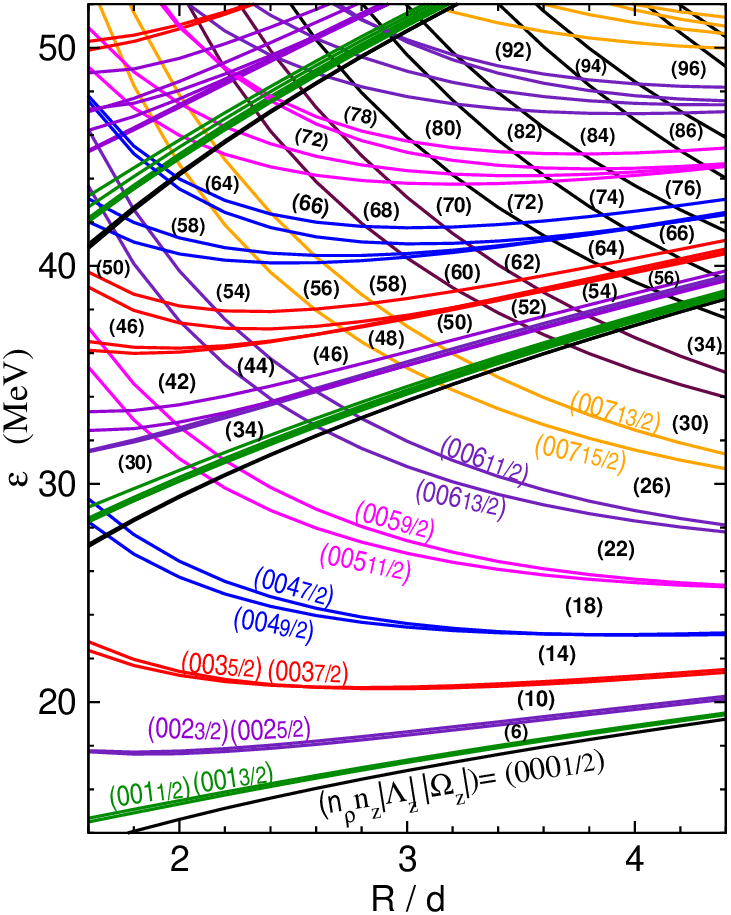} 
\caption{(Color online) Single-particle energies for a toroidal
  nucleus as a function of the aspect ratio $R/d$.  The states are
  labeled by quantum numbers $(n_\rho, n_z, |\Lambda_z|,|\Omega_z|)$
  which are displayed only for the lowest states.  The locations of
  the toroidal shell are shown as bracketed numbers, $(N)$, at their
  corresponding toroidal deformations, $R/d$.  }
\label{fig2}
\end{figure}

\begin{enumerate}
\item
As a function of increasing toroidal deformation $R/d$,  there are sequences of toroidal shells
occurring at an interval of $\Delta (R/d)$$\sim$0.3 and $\Delta N$=2.

\item
For a fixed value of toroidal deformation $R/d$,  there are sequences of toroidal shells 
occurring at an interval of $\Delta N$=10 arising from nucleons occupying  four $(n_\rho$+$n_z$=1, $\Lambda_z)$ states  and a single $(n_\rho+n_z$=0,$\Lambda_z')$ state.

\end{enumerate}

  There are toroidal shells with nucleon numbers (30), (42), (46), and
  (50) at $R/d$$\sim$1.8, (34+10$m$) and (58) at $R/d$$\sim$2.2,
  (46+10$m$) and (72) at $R/d$$\sim$2.4, (48+10$m$) at $R/d$$\sim$2.9,
  (50+10$m$) at $R/d$$\sim$3.2, (52+10$m$) at $R/d$$\sim$3.5,
  (54+10$m$) at $R/d$$\sim$3.8, and (56+10$m$) at $R/d$$\sim$4.2.  The
  integer $m$ in these toroidal shell series starts with $m$=0 and
  terminates with $m$=3, or 4, until the $n$=$n_\rho$+$n_z$=2
  single-particle states are reached.

 Nuclei with toroidal shells gain extra stability at the associated
 toroidal deformation.  Consequently, they present themselves as good
 candidates in the search for excited toroidal local energy minima in
 microscopic model calculations, as carried out in previous Strutinsky
 shell correction calculations or in self-consistent men-field
 calculations \cite{Won72,Won73,Zha10,Cao18}.

\begin{table}[h]
\vspace*{0.3cm}
  \caption{ Locations of toroidal shells in the
    intermediate mass region.  }

\vspace*{0.3cm}
\centering

\begin{tabular}{|c | c |}
\hline
Toroidal & Toroidal shell nucleon numbers \\ 
 deformation \!$R/d$ & $(N)$ \\ \hline
$\sim$1.8 & (30), (42), (46), (50) \\
$\sim$2.2 & (34), (44), (54), (58),(64) \\
$\sim$2.5 & (46), (56), (66), (72) \\
$\sim$2.9& (48), (58), (68), (78)\\
$\sim$3.2& (50), (60), (70), (80)\\
$\sim$3.5& (52), (62), (72), (82), (92) \\
$\sim$3.8& (54), (64), (74), (84), (94)\\
$\sim$4.2& (56), (66), (76), (86), (96)\\
\hline
\end{tabular}
\label{tab1}
\end{table}

It is interesting to note that in the intermediate mass region, many
toroidal shells of different nucleon numbers occur at approximately 
the same toroidal deformation.  This feature facilitates
the combinations of different neutron and proton numbers to maximize
the shell effects at the same toroidal deformation and at the same
time minimize the instability against beta decay. However, because of the Coulomb interaction, the proton toroidal shell nucleon
number and toroidal deformations will be slightly modified from those
listed in Table \ref{tab1}.  The repulsive single-particle proton
potential is greatest near the region of the greatest nuclear
densities and therefore the single-particle Coulomb potential behaves
approximately as an inverted harmonic oscillator in the meridian
plane.  To the lowest order of modification, it will effectively
modify the harmonic oscillator frequency $\hbar \omega_\perp$ so that one effectively
speaks of $\hbar \omega_\perp(\rm neutron)$ for the neutron and a
slightly modified $\hbar \omega_\perp(\rm proton)$ for the proton.  We
expect from Eq.\ (\ref{eq6}) that similar to the neutron toroidal
shells there will likewise be proton toroidal shells at various
toroidal deformations.  Furthermore, because of the regularity of the shell structure and 
the frequent occurrences of the toroidal shells in both nucleon numbers
and toroidal deformations one expects that there will be many
combinations of protons and neutron toroidal shells occurring at the
same toroidal deformation to make them favorable for the stabilization
due to the nuclear shell effects. 
 Future microscopic models will
provide a better description of the toroidal nuclei possibilities
in the intermediate mass region.

\section{ Toroidal High-Spin Isomers in the Intermediate Mass Region}

In addition to the toroidal shell structure characterized by the
nucleon number $N$ at the toroidal deformation $R/d$, there is the
spin degree of freedom that is worth exploring for the intermediate
mass region.  Relative to an even-even core of a toroidal nucleus
occupying the lowest toroidal single-particle states at a given
toroidal deformation $R/d$, toroidal high-spin isomers with different
spins as yrast states may be constructed with the spin-aligning
Bohr-Mottelson particle-hole excitations \cite{Boh81} as carried out
in \cite{Ich12,Ich14,Sta14,Sta15,Sta15a,Sta16,Kos17,Sta17,Kos18}.  The $I$=0
toroidal core nucleus may be in a local energy minimum, if the nucleus
has both neutron and proton toroidal shells and the associated shell
correction energy is strong enough to allow an energy minimum
\cite{Cao18}.  On the other hand, if the shell correction is not
strong enough to allow an energy minimum, or, if the nucleon numbers
at that toroidal deformation fall into regions of high single-particle
state density with positive shell corrections, then the toroidal
nucleus will be away from a local energy minimum
\cite{Kos17,Sta17,Kos18}.  In either case, high-spin toroidal isomers
can nevertheless be constructed from the $I$=0 toroidal core using the
spin-aligning Bohr-Mottelson particle-hole excitations \cite{Boh81} by
promoting nucleons occupying states with an angular momentum opposite
to a chosen symmetry axis to occupy empty states with an angular
momentum along the symmetry axis.

Using the cranking frequency $\hbar \omega$ as a Lagrange multiplier,
the particle-hole excitation leading to a particular angular momentum
yrast $I$=$I_z$ state can be obtained by occupying the lowest Routhian
single-particle orbitals as a function of $\hbar \omega$, as was shown
in previous cranked self-consistent Hartree-Fock calculations
\cite{Ich12,Ich14,Sta14,Sta15,Sta15a,Sta16,Kos17,Sta17,Kos18}.  The
single-particle Routhian $\epsilon_{\rm Routhian}$$(n_\rho n_z
\Lambda_z \Omega_z)$, under the constraint of the non-collective
aligned angular momentum is related to the single-particle energy
$\epsilon$$(n_\rho n_z \Lambda_z \Omega_z)$ by
\begin{equation}
\epsilon_{\rm Routhian}( n_\rho n_z  \Lambda_z \Omega_z)
=\epsilon( n_\rho  n_z \Lambda_z \Omega_z)
-\hbar \omega \Omega_z.
\end{equation}
It is necessary to investigate the shell structure in the space of
$N$, $R/d$, and $\hbar \omega$.  
As an example,
we shall show the shell structure at
$R/d$=2.9 where many toroidal shells are located.  It is
of interest to see how different toroidal shell structure evolve as a
function of the cranking frequency $\hbar \omega$.  We show in
Fig.\ \ref{fig3} the multi-dimensional nature of the shell structure
by exhibiting the single-particle state energy as a function of $R/d$
in Fig. \ref{fig3}(a), and the the single-particle Routhian energy as
a function of $\hbar \omega$ at $R/d$=2.9 in Fig. \ref{fig3}(b).  We
can use Fig.~\ref{fig3}(b) to determine the spin value, $I$=$I_z$, as
a function of $N$ and $\hbar \omega$ at $R/d$=2.9.  Specifically, for
a given $N$ and $\hbar \omega$, the aligned $I$=$I_z$ from the $N$
nucleons occupying the lowest Routhian energy states can be obtained
by summing $\Omega_{zi}$ over all states below the Fermi energy and
the summed aligned angular momentum $I$ is a step-wise function of the
Lagrange multiplier $\hbar \omega$ \cite{Ring80}, with each $I$
spanning a small region of $\hbar \omega$.  By such a construction, we
obtain selective regions of low Routhian energy level densities at
different $(N,I,R/d)$ configurations at the top of the Fermi energy and
the toroidal deformation $R/d$.  These $(N,I,R/d)$ configurations are
favorable candidates in search of toroidal high-spin isomers
\cite{Sta14} in microscopic models.  We show the locations of
favorable $(N,I)$ configurations for $R/d$=2.9 in Fig.\ {\ref{fig3}
  (b).
\begin{figure} [t]
\centering \includegraphics[scale=0.43]{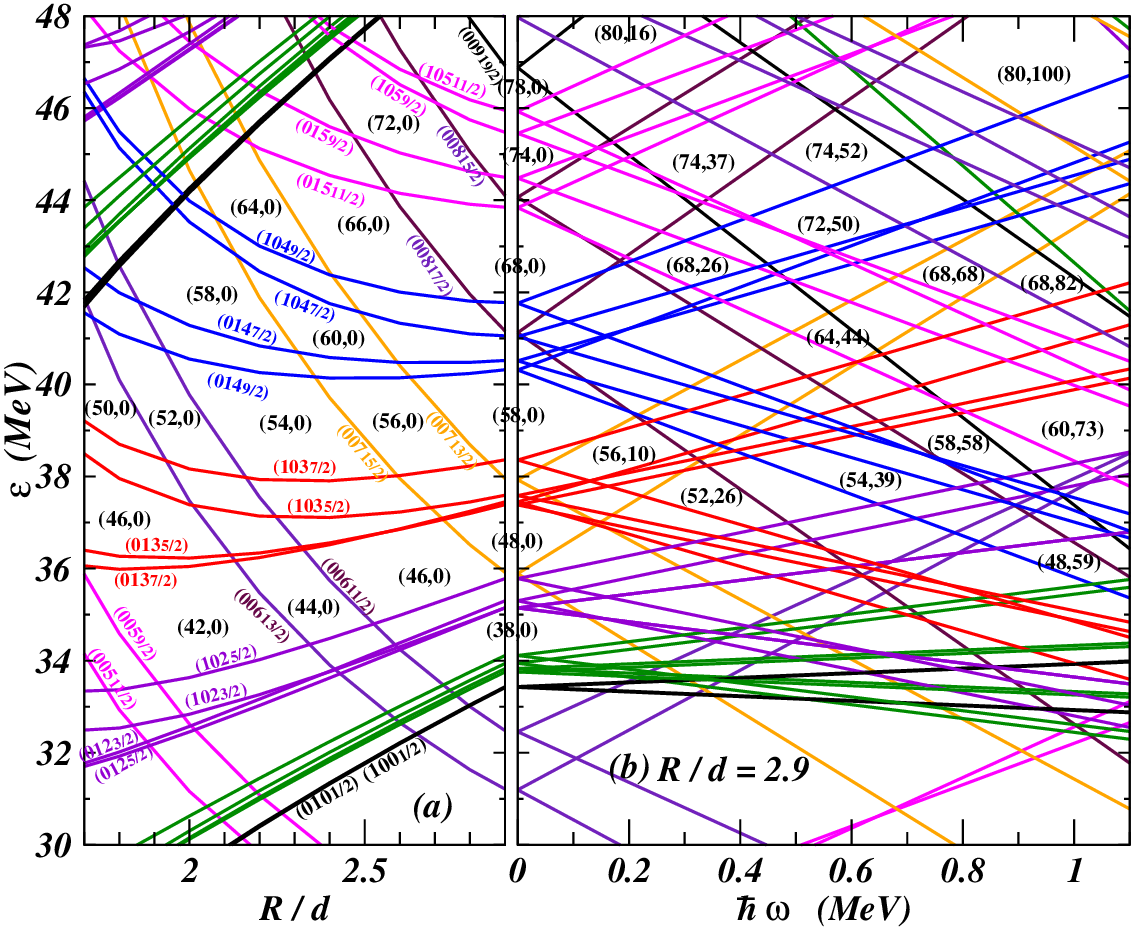}
\caption{(Color online) (a) Single-particle state
  energy $\epsilon$ for a toroidal nucleus as a function of the aspect
  ratio $R/d$.  The states are labeled with quantum numbers
  $(n_\rho,n_z |\Lambda_z|,|\Omega_z|)$.  (b)
  Single-particle Routhian $\epsilon_{\rm Routhian}$ as a function of
  the cranking frequency $\hbar \omega$, at the toroidal deformation
  $R/d=2.9$.  The bracketed numbers are favorable $(N,I)$
  configurations at this deformation.  }
\label{fig3}
\end{figure}

It is interesting to note in Fig.\ \ref{fig3}(b) that at $R/d$=2.9 and
$\hbar \omega$$\sim$0.4 MeV, the shell regions of low Routhian energy
level density occur for $(N,I)$=(52,26), (68,26),  and (74,37)
 which may allow various $(Z,I_{\rm proton})$ and
$(N,I_{\rm neutron})$ combinations to be good candidates for toroidal
high-spin isomers.  At $\hbar \omega$$\sim$0.6 MeV, the shell regions
of low Routhian density occur for (54,39), (64,44), (72,50), and
(74,52).  At $\hbar \omega$$\sim$0.8 MeV, the shell regions of low
Routhian density occur for  (58,58) and  (68,68), and at $\hbar
\omega$$\sim$1.0 MeV they occur at (48,59), (68,73), (68,82), and
(80,100).  Note that there are many nucleon numbers (such as $N$=52, 74, 80, 54, 64,...)
in these favorable configurations 
 that are not toroidal shell numbers associated with extra  stability at $\hbar \omega$=0
 without cranking at $R/d$=2.9.   These nucleon numbers  nevertheless may 
 be favorable for  high-spin isomers because of their
 shell structure as a function of $\hbar \omega$ under a cranking motion, 
 as in the analogous case in the superheavy nuclei region \cite{Sta17}.
 From this viewpoint, it is not always necessary to have a toroidal core located at an energy minimum to
 make the Bohr-Mottelson spin-aligning single-particle excitations for toroidal high-spin isomers.
 What is necessary is  the low Routhian energy density under the cranking motion that provides 
the  favorable condition to stabilize toroidal high-spin isomers.   
Favorable $(N,I)$ configurations occur at other toroidal
deformations $R/d$ as well.  

The results in Figs. \ref{fig3} demonstrate that just as it is with
light mass nuclei, toroidal high-spin isomers are also expected in the
intermediate mass region.  The shell structure is however a very
complicated function of toroidal deformation and the cranked
frequency.  Thus the occurrence of the favorable combination shell
region $(N,I, R/d)$ for toroidal isomers can only be treated on a case
by case basis.  The search for high-spin isomers can proceed
as in \cite{Zha10,Sta14,Sta14,Sta15,Sta15a,Sta16,Kos17,Sta17,Kos18} by using figures similar to
Fig.\ \ref{fig3} as a guide.

The spin and the excitation energy of a toroidal high-spin isomer can be calculated
in the self-consistent mean-field theory.  It can also be estimated in
the toroidal shell model as was carried out in \cite{Cao18}.  In such
an estimate, upon promoting a nucleon from an initial orbital
$(n_\rho, n_z, \Lambda_z,\Omega_z)_i$ to become a hole state and to
occupy an empty final orbital $(n_\rho, n_z,\Lambda_z,\Omega_z)_f$,
one calculated the change in the aligned angular momentum and the
excitation energy for such a promotion from the quantum numbers and
the single-particle energies of the particle-hole states.  The total
spin $I$=$I_z$ and total exudation energy $E_I$$-$$E_{I=0}$ are then
the sums from all particle-hole promotions.  The spin of the toroidal
nucleus is given by
\begin{eqnarray}
I_z=\sum_{f}^{\rm particle~ states}  \Omega_{zf} - \sum_{i}^{\rm hole ~states}  \Omega_{zi},
\end{eqnarray}
and the total excitation energy is given by
\begin{eqnarray}
E_I-E_{I=0}&=&\sum_{f}^{\rm particle~ states} \epsilon(n_{z} n_{\rho } \Lambda_{z }
 \Omega_{z} )_f
 \nonumber\\
 & &
-\sum_{i}^{\rm hole~states} \epsilon(n_{z} n_{\rho} \Lambda_{z} \Omega_{z} )_i.
\end{eqnarray}

It should however be stressed that what has been presented here with
the analytical toroidal single-particle shell model provides only an
intuitive guide on the interesting nuclei where toroidal 
high-spin isomers may be searched and located.  Whether these states
turn out to be local energy minima will need to rely on reliable
microscopic models such as the non-relativistic mean-field or
relativistic mean-field calculations as carried out in
\cite{War07,Vin08,Sta09,Zha10,Ich12,Ich14,Sta14,Sta15,Sta15a,Sta16,Men13,Ren17,Zhao10,Lal05,Kos17,Sta17,Kos18,Cao18,Afa18}.

\section{{Toroidal nuclei with vorticities}\label{sec4}}

The geometrical shape of a toroidal nucleus provides a natural way to
describe nuclear vorticities as shown in Fig.\ 10 of \cite{Won73} and
Fig.\ \ref{fig1}.  The concept of vortex nucleus is best examined in
the limit of a toroidal nucleus with a large major radius $R$ so that
we can neglect $d/R$ as well as  
 the
difference between $\hbar \omega_\perp$ and $\hbar \omega_\perp '$ are small
and can be neglected.  Within this approximation, we can re-write
Eqs. (\ref{eqr}) and (\ref{eqz}) as
 \begin{eqnarray}
&&\biggl [  -\frac{\hbar ^2}{2m} ( \frac{\partial^2}{\partial  q ^{2}}
- \frac{\Lambda_z^2-\frac{1}{4}}{R^2}) 
 -\frac{\hbar ^2}{2m} \frac{\partial^2}{\partial  z ^{2}}
 +\frac{1}{2} m \omega_\perp^2 q^{2}+\frac{1}{2} m \omega_\perp^2 z^{2}
 \nonumber\\
&&
~- \! \frac{2\kappa \hbar}{m \oo \omega}
\frac{m \omega_\perp^2 }{R/d}
s_z \Lambda_z\! -\! \epsilon({n_\rho n_z \Lambda_z \Omega_z})\biggr ]{{\cal R}_{n_\rho}(\rho)Z_{n_z}(z)}\!=\!0.~~~
\label{eq14}
\end{eqnarray}
where $q=\rho-R$.  We can transform the 2-dimensional $(q,z)$
coordinates to polar coordinates $(r_\perp, \theta)$ where
\begin{eqnarray}
r_\perp=\sqrt{q^2+z^2}, ~~~  \theta=\tan^{-1}(z/q),
\end{eqnarray}
and $\theta$ is the poloidal angle  as shown in Fig.\ \ref{fig1}.  We can rewrite
Eq. (\ref{eq14}) as
 \begin{eqnarray}
&&\biggl [  -\frac{\hbar ^2}{2m} \left (\frac{1}{r_\perp} \frac{\partial}{\partial r_\perp}
r_\perp \frac{\partial}{\partial r_\perp}- \frac{\Lambda_\perp^2}{r_\perp^2} \right )
\!+\!\frac{1}{2} m \omega_\perp^2 r_\perp^{2}\!+\! \frac{\hbar^2(\Lambda_z^2-\frac{1}{4})}{2mR^2}
\nonumber\\
&&-  \frac{2\kappa \hbar}{m \oo \omega}
\frac{m \omega_\perp^2 }{R/d}
s_z \Lambda_z\! - \! \epsilon({n_\perp \Lambda_\perp \Lambda_z \Omega_z})\! \biggr ]\!{\Re}_{n_\perp\! \Lambda_\perp}\!(r_\perp)e^{i\Lambda_\perp\theta}  \!\!\!\!=\!0.
\label{eq27}
\end{eqnarray}
In Eq.\ (\ref{eq14}), the two-dimensional harmonic oscillator
nucleon wave function ${{\cal R}_{n_\rho} (\rho)Z_{n_z}(z)}$ in
($q$,$z$) coordinates in the meridian plane with quantum numbers
$(n_\rho, n_z)$ has been transformed into ${\Re}_{n_\rho
  |\Lambda_\perp|} (r_\perp)e^{i\Lambda_\perp\theta}$ in $(r_\perp,
\theta)$ coordinates with quantum numbers $(n_\perp, \Lambda_\perp)$ in the above equation.
Note that the wave function $e^{i\Lambda_\perp\theta}$ describes the
state with vorticity $\Lambda_\perp$ associated with a circulating
vortex current around the poloidal angle $\theta$ in the meridian
plane as shown in Fig.\ 1.  It is the poloidal angular momentum along the poloidal angular
direction.  If only a single $e^{i\Lambda_\perp\theta}$ state is
occupied, we have a state of vorticity $\Lambda_\perp$, but if 
there is a pair of nucleons occupying both $\pm|\Lambda_\perp|$
states, then the vorticities of these two nucleons cancel each other,
and we have a state of zero total vorticity.  If by construction when there
is a bias in occupying more positive $\Lambda_\perp$ states than
negative $\Lambda_\perp$ states (or vice versa), then there will be a net non-zero
$\Lambda_\perp$ and consequently a net non-zero total vorticity.

The two-dimensional harmonic oscillator energy in $(n_\rho n_z)$ and
in $(n_\perp \Lambda_\perp)$ are related by
\begin{eqnarray}
(n_z+n_\rho+1)\hbar \omega_\perp  = (2 n_\perp +  |\Lambda_\perp|+1)\hbar  \omega_\perp .
\end{eqnarray}
We have the equivalence between the toroidal states in $(n_z n_\rho)$
and in $(n_\perp \Lambda_\perp)$ in Table \ref{tab2} which shows that
a nucleon in a toroidal nucleus residing at states with
$n_z$+$n_\rho$$\ge$1 possesses a non-zero vorticity quantum number
$\Lambda_\perp$.  We can re-label the set of states with quantum
numbers $(n_\rho n_z)$ to $(n_\perp \Lambda)$,
\begin{eqnarray}
|n_\rho n_z \Lambda_z \Omega_z\rangle  \to |n_\perp \Lambda_\perp
\Lambda_z \Omega_z\rangle.
\end{eqnarray}
Clearly, the $\Lambda_\perp=\pm |\Lambda_\perp|$ states are
degenerate.  Vorticity is a new degree of freedom not available for
light nuclei studied in
\cite{Zha10,Ich12,Ich14,Sta14,Sta15,Sta15a,Sta16,Cao18}.
\begin{table}[h]
\vspace*{0.3cm}
  \caption{ The equivalence of $(n_\rho,n_z) \to (n_\perp,\Lambda_\perp)$
for the toroidal single-particle states.
  }
\vspace*{0.3cm}
\centering
\begin{tabular}{|c | c |c | c|}
\hline
$n_\rho+n_z$\!& $(n_\rho n_z)$ states    & $(n_\perp,\Lambda_\perp)$ states &Number \\ 
  &  &  & of states\\  \hline
0 & (0, 0) &  (0,0) & 1 \\ \hline
1 & (0,1) (1,0) & (0,+1) (0,-1) & 2 \\ \hline
2 & (0,2),(2,0),(1,1) & (0,+2),(0,-2),(1,0)& 3 \\ \hline
3 &\!(0,3),(3,0),(1,2),(2,1)\!&(0,3),(0,-3),(1,1),(1,-1)\!& 4 \\ 
\hline
\end{tabular}
\label{tab2}
\end{table}

The single-particle energy of $|n_\perp \Lambda_\perp \Lambda_z
\Omega_z\rangle$ is therefore given from Eq.\ (\ref{eq27}) by
\begin{eqnarray}
\epsilon({n_\perp \Lambda_\perp \Lambda_z \Omega_z})& &=(2n_\perp+|\Lambda_\perp|+1) \hbar \omega_\perp
+ \frac{\hbar^2(\Lambda_z^2-1/4)}{2 m R^2} 
\nonumber\\
& &- \frac{2\kappa \hbar}{m{\oo \omega}}
\frac{m \omega_\perp^2 }{R/d}s_z \Lambda_z.
\label{eq29}
\end{eqnarray}

The vorticity of a nucleus or a nucleon is a quantized quantity and is
measured in units of $\hbar$.  After the symmetry $z$-axis has been
chosen, as for example as the axis pointing in the upward direction in
Fig.\ \ref{fig1}, then we can designate the vorticity to be positive
by the right-handedness of the poloidal flow on the right rim of the
nucleus, according to the sense of the poloidal angle $\theta$ as
shown in Fig.\ \ref{fig1} between the $\rho$ axis and the radial
vector $\bf r_\perp$.  Negative vorticity corresponds to the case of
left-handedness of the poloidal flow.  The handedness property of the
poloidal flows allows one to associate the vorticity quantum numbers
also with equivalent chirality quantum numbers, with positive and
negative vorticities associated with positive and negative chiralities
respectively.  Clearly, if the toroidal vortex nucleus has a non-zero spin
about the symmetry axis, the toroidal vortex nucleus with opposite
vorticities (or chiralities) are distinguishable physical states.

Vorticity is a good quantum number and it is a conserved quantity.  It
would be of interest to study the production the decay of these nuclei
to see if they may show up as exotic meta-stable states with a toroidal
topology and flow patterns.

\section{How to construct a vortex nucleus}

We can use Table \ref{tab2} to construct a nucleus with a total
vorticity $\Lambda_\perp^{\rm total}$.  The vorticity quantum number
of a nucleon $\Lambda_\perp$ can be both positive and negative and the
total vorticity $\Lambda_\perp^{\rm total}$ for an $N$ (or $Z$)
nucleon system is the sum of the vorticities of its constituent
nucleons, $\Lambda_{\perp i }$,
\begin{eqnarray}
\Lambda_\perp^{\rm total}=\sum_i^{N~({\rm or}~Z)} \Lambda_{\perp i}.
\end{eqnarray}

We study a concrete example and focus our attention on the toroidal
nucleus with $Z$=48 and $N$=58 at $R/d$=2.9 whose large
single-particle energy gaps in Fig.\ \ref{fig2} make it likely to have an
excited toroidal energy minimum stable against expansion and
contraction of the major radius at that toroidal deformation.  We
examine first the vorticities and the energy for protons with $Z$=48
in Tables \ref{tab3}, \ref{tab4}, and Fig.\ 2.  By populating the lowest
single-particle states at this possible toroidal energy minimum, the 8
topmost occupied states are $(n_\perp, \Lambda_\perp,
|\Lambda_z|,\Omega_z)$=(0,$\pm$1,2,($\pm$3/2)) and
(0,$\pm$1,2,($\pm$5/2)), each of which is degenerate with
$\Lambda_z=\pm | \Lambda_z|$, $\Omega_z$=$\pm$$|\Omega_z|$.  So, we
have $\epsilon(n_\perp, \Lambda_\perp,
|\Lambda_z|,\Omega_z)$=$\epsilon$(0,1,2,(+3/2))=$\epsilon$(0,1,2,(-3/2)),
and $\epsilon(n_\perp,
\Lambda_\perp,|\Lambda_z|,\Omega_z)$=$\epsilon$(0,1,2,(+5/2))=$\epsilon$(0,1,2,(-5/2)).
Each of these four states,
$(n_\perp,|\Lambda_\perp|,|\Lambda_z|,\Omega_z)$=$\{$ (0,1,2,(+3/2)),
~(0,1,2,(-3/2)), ~(0,1,2,(+5/2)), ~(0,1,2,(-5/2)) are doubly
degenerate with $\Lambda_\perp=\pm | \Lambda_\perp|$.  For the
configuration in which the 8 topmost occupied state for $Z$=48 are
given in Table \ref{tab3}, the total vorticity $\Lambda_\perp^{\rm
  total}$ and the total spin about the $z$-axis $\Omega_z^{\rm total}$
are zero.

\vspace*{-0.6cm}
\begin{table}[H]
\vspace*{0.4cm}
\caption{ The configurations of the 8 topmost  occupied single-particle
  states at the top of the Fermi energy
  for $Z$=48 and $R/d$=2.9 in Fig.\ 2.  In this configuration, the total
  proton vorticity for this state, $\Lambda_{\perp{\rm proton}}^{\rm total}$
  and the total spin   $\Omega_z^{\rm total}$   are zero.  }
\vspace*{0.2cm} 
\centering
\begin{tabular}{|c | c | c |c |}
\hline
     $n_\perp$  &  Vorticity    $\Lambda_\perp$  & $ |\Lambda_z|$ &$ \Omega_z$ \\ \hline
   0 &  ~1  & 2  & ~3/2 \\ \hline
   0 &  $-$1  & 2  & ~3/2 \\ \hline
   0 &  ~1  & 2  & $-$3/2 \\ \hline
   0 &  $-$1  & 2  & $-$3/2 \\ \hline
   0 &  ~1  & 2  & ~5/2 \\ \hline
   0 &  $-$1  & 2  & ~5/2 \\ \hline
   0 &  ~1  & 2  & $-$5/2 \\ \hline
   0 &  $-$1  & 2  & $-$5/2 \\ \hline
\end{tabular}
\label{tab3}
\end{table}

Now suppose we promote all four proton states in Table \ref{tab3} 
with $\Lambda_\perp$=$-1$ from
such $\Lambda_\perp$=$-$1 states to the next un-occupied level with
$\Lambda_\perp$=+1 by particle-hole excitations. According to Fig.\ 2, the next un-occupied  levels are
$(n_\perp, \Lambda_\perp, |\Lambda_z|,\Omega_z)$=$(0,\Lambda_\perp$=+
1,$|\Lambda_z|$=3,$|\Omega_z|$=5/2 \& 7/2).  We obtain the set of
occupied states in Table \ref{tab4}, with the promoted particle
states in bracketed numbers.
\begin{table}[h]
  \caption{ The configurations of the 8 topmost occupied states for
    $Z$=48 at $R/d$=2.9 in Fig.\ 2 after the vortex-creating 4p-4h excitations
    with the promoted particle configurations shown in bracketed
    numbers.  In this configuration, the total proton vorticity is
    $\Lambda_{\perp {\rm proton}}^{\rm total}=8\hbar$ and
    $\Omega_{z~{\rm proton}}^{\rm total}$=0.  }
\vspace*{0.2cm}
\centering
\begin{tabular}{|c | c | c |c |}
\hline
     $n_\perp$  &  Vorticity    $\Lambda_\perp$  & $ |\Lambda_z|$ &$ \Omega_z$ \\ \hline
   0 &  ~1  & 2  & ~3/2 \\ \hline
   0 &  {\color{red}~(1)} &  {\color{red} (3)}  & {\color{red}~(5/2)} \\ \hline
   0 &  ~1  & 2  & $-$3/2 \\ \hline
   0 &  {\color{red}~(1)}   &  {\color{red} (3)}     &  {\color{red} ($-$5/2)} \\ \hline
   0 &  ~1  & 2  & ~5/2 \\ \hline
   0 &  {\color{red}~(1)}   &  {\color{red} (3)}     &   {\color{red}~(7/2)} \\ \hline
   0 &  ~1  & 2  & $-$5/2 \\ \hline
   0 &  {\color{red}~(1)}  &   {\color{red} (3)}    &  {\color{red} ($-$7/2)}  \\ \hline
\end{tabular}
\label{tab4}
\end{table}

The excitation energy of this 4p-4h state with 8 units of vorticity
from Tables \ref{tab3} and \ref{tab4} in promoting the set of $\{i\}$
nucleons from the $|n_\perp \Lambda_\perp\Lambda_z \Omega_z\rangle$
states to the $|n_\perp '\Lambda_\perp'\Lambda_z' \Omega_z'\rangle$
states is
\begin{eqnarray}
 E_{\Lambda_\perp=8  }^{\rm proton}\!\!\!-
E_{\Lambda_\perp=0}^{\rm proton}
&&=\sum_{{\rm all}~i}\left ( \epsilon_{n_\perp' \Lambda_\perp' \Lambda_z' \Omega_z'} (i)
-  \epsilon_{n_\perp \Lambda_\perp \Lambda_z \Omega_z} (i)\right ).~~~~~
\nonumber\\
&& =~~~[\epsilon(0,1,3,5/2)+\epsilon(0,1,3,-5/2)
\nonumber\\
&&~~~+\epsilon(0,1,3,7/2)+\epsilon(0,1,3,-7/2)]
\nonumber\\
&&~~ -  [\epsilon(0,1,2,3/2)+\epsilon(0,1,2,-3/2)
\nonumber\\
&&~~~+\epsilon(0,1,2,5/2)+\epsilon(0,1,2,-5/2)],
\end{eqnarray}
which can be obtained from the eigenvalue equation Eq.\ (\ref{eq29}).
Assuming negligible spin-orbit interaction, which is probably small,
we have from the protons with $Z$=48 at $R/d$=2.9 with a total proton vorticity 
$\Lambda_{\perp {\rm proton}}^{\rm total}$= 8$\hbar$, 
\begin{eqnarray}
E_{\Lambda_\perp=8}^{\rm proton}\!\!-
E_{\Lambda_\perp=0}^{\rm proton}&=&4\times \frac{\hbar^2(\Lambda_z'^2-\Lambda_z^2)}{2 m R^2}
\nonumber\\
&=&4\times \frac{\hbar^2(3^2-2^2)}{2 m R^2}=\frac{20\hbar^2}{2 m R^2}.
\label{eq37}
\end{eqnarray}

We turn now to the vorticities and the energy for neutrons with $N$=58
at $R/d$=2.9 in Fig.\ 2.  Procedures similar to those given above lead to Table
\ref{tab5}, a total neutron vorticity $\Lambda_{\perp {\rm
    proton}}^{\rm total}$= 8$\hbar$, and an excitation energy
\begin{eqnarray}
E_{\Lambda_\perp=8  }^{\rm neutron}\!\!-
E_{\Lambda_\perp=0}^{\rm neutron}&=&4\times \frac{\hbar^2(\Lambda_z'^2-\Lambda_z^2)}{2 m R^2}
\nonumber\\
&=&4\times \frac{\hbar^2(4^2-3^2)}{2 m R^2}=\frac{28\hbar^2}{2 m R^2}.
\label{eq38}
\end{eqnarray}
\begin{table}[h]
\vspace*{-0.3cm}
  \caption{ The configurations of the 8 topmost states with different
    state quantum numbers at the top of the Fermi energy for $N$=58
    and $R/d$=2.9 in Fig 2, after the vortex-creating 4p-4h excitation
    with the particle configuration shown in bracketed numbers.  The
    total vorticity for this $N=58$ state is $\Lambda_{\perp{\rm
        neutron}}^{\rm total}=8\hbar$ and $\Omega_{z{\rm~
        neutron}}^{\rm total}$=0. .  }  \centering
\vspace*{0.2cm}
\begin{tabular}{|c | c | c |c |}
\hline
     $n_\perp$  &  Vorticity    $\Lambda_\perp$  & $ |\Lambda_z|$ &$ \Omega_z$ \\ \hline
   0 &  ~1  & 3  & ~5/2 \\ \hline
   0 &  {\color{red}~(1)} &  {\color{red} (4)}  & {\color{red}~(7/2)} \\ \hline
   0 &  ~1  & 3  & $-$5/2 \\ \hline
   0 &  {\color{red}~(1)}   &  {\color{red} (4)}     &  {\color{red} $(-$7/2)} \\ \hline
   0 &  ~1  & 3  & ~7/2 \\ \hline
   0 &  {\color{red}~(1)}   &  {\color{red} (4)}     &   {\color{red}~(9/2)} \\ \hline
   0 &  ~1  & 3  & $-$7/2 \\ \hline
   0 &  {\color{red}~(1)}  &   {\color{red} (4)}    &  {\color{red} $(-$9/2)}  \\ \hline
\end{tabular}
\label{tab5}<
\end{table}

The total excitation energy of $(Z,N)=(48,58)$ with a total vorticity of 16$\hbar$ is
\begin{eqnarray}
E_{\Lambda_\perp=16}-E_{\Lambda_\perp=0}=\frac{48\hbar^2}{2 m R^2},
\end{eqnarray}
which decreases as $R/d$ increases.  Here the number 4 on the
right-hand sides of Eqs.\ (\ref{eq37}) and (\ref{eq38}) is the number
$N_{ph}$ of particle-hole excitations that promote the negative
$\Lambda_\perp$=$-1$ to the positive $\Lambda_\perp$=+1 states (or
vice-versa) and can be changed to get a state with a different
vorticity.  Thus, the greater the number of particle-hole promotion
$N_{ph}$, the greater is the vorticity, and the greater is the effect
to lowering the energy of the system by moving to greater $R$ values.
In other words, there is an energy associated with a given vorticity,
and for a given vorticity, the larger the $R$, the lower the energy of
the system.  The toroidal system likes to expand to a greater radius
$R$, if the nucleus has a vorticity.

In the above example, we have made a 4p-4h excitation to obtain an
excited state of vorticity $\Lambda_\perp($neutron$)$=
$\Lambda_\perp($proton$)$=8$\hbar$.  It is possible
to make $n$p$-$$n$h excitations with $n$=1 to 4 for both neutrons and
protons, with the only modification that the total $\Omega_z$ would
vary according to the $\Omega_z$ value of the particle-hole
configuration.  The signature of the vorticity occurrence is therefore
a set of states with vorticity $\Lambda_\perp$=2,4,6,8,.. and 16, with
about equal energy spacing.

What we have presented is an example of how we can construct a
toroidal vortex nuclei with different vorticities for $Z$=48 and
$N$=58 at $R/d$=2.9.  Other toroidal vortex nucleus can be similarly
constructed for different neutron and proton numbers at various toroidal deformations.  The large
number of toroidal shells as listed in Table \ref{tab1} provide a
large pool of neutron and proton combinations for which toroidal
vortex nuclei may be constructed at different toroidal deformations.
Therefore, there can be many different toroidal vortex nuclei with
different neutron and proton numbers constructed by making
vortex-creating particle-hole single particle excitations in the
intermediate mass region.  The possibility for nucleons possessing a
non-zero vortex quantum number with $n$=$n_\rho$+$n_z$$>$1 in the
heavy mass region implies that toroidal vortex nuclei are also
expected in the heavy mass region.

\section {Possible mechanisms for the production of toroidal high-spin isomers and toroidal vortex isomers}

The method of production and detection of toroidal nuclei depends on
their lifetimes which are however difficult to estimate because they
require the knowledge of the potential energy surface and the effective mass
parameter in many different degrees of freedom. Many mean-field
calculations from independent groups have been carried out
\cite{Ich12,Ich14,Sta14,Sta15,Sta15a,Sta16,Kos17,Sta17,Kos18,Cao18}. They
indicate stability against small amplitude oscillations, as
3-dimensional calculations with 3-dimensional noises have been carried
out and they yield toroidal high-spin isomeric energy minima. However,
the stability against large amplitude oscillations, as for example
against the Plateau-Raleigh instability \cite{plateau,rayleigh14,eggers97}
 is not known.  Subject to further studies to confirm or  refute the experimental 
 evidence for possible population of toroidal high-spin isomers in \cite{Cao18}, the tentative
 extraction of the width for toroidal $^{28}$Si high-spin
isomer indicates that the widths may be  broad for low-spin states,
become narrower than the instrument bin size of a 3-4 MeV for the
I=28, 36, and 44 states, and they become broader again for possible
higher-spin states.  We may expect that
the lifetime depends on the spins and the excitation energies.  For
low-spin states, as they lie in highly excited region of states of
similar angular momentum from the co-existing sphere-like geometry,
there may be extensive mixing that will likely broaden the widths and
shorten the lifetimes of these low-spin states.  Longer lifetimes may
be reached by promoting nucleons to populate higher but well-bound
single-particle orbitals to lead to higher-spin toroidal high-spin
isomers as yrast states with hardly any mixing of states of similar
angular momentum from the sphere-like geometry.  However, when the
nucleons are promoted to higher-energy orbitals that may be above or
near the nucleon drip line in higher excitation energies, the widths
of these high-spin states may be broadened again, with shorter
lifetimes.

\subsection{Production of light-mass toroidal isomers  by Elastic Scattering}

In recent years, TPCs (Time Projection Chambers) have been used to study the nuclear
spectroscopy of meta-stable nuclei
\cite{Ahn12,Rog11,Ket13,Ayy17,ATTPC17}. The idea is to use a chamber
of noble gas under a high voltage so that the gas itself or an
embedded solid layer serves as the target, and the nuclear
trajectories show up as tracks. The production of a composite nucleus
with a long half-life would show up as a single track with the mass
and charge arising from the fusion of the projectile and target
nuclei.  The production of binary products indicates a two$\to$two
reaction from which one can examine the elastic and inelastic channels
and study the excitation function and angular distribution to search
for various meta-stable states. Previously, many meta-stable states
formed by colliding various projectile nuclei with an active He target
have been found by such a technique \cite{Ahn12}.  We can search for
toroidal high-spin isomers as resonances or meta-stable nuclei by
bombarding projectile nucleus on an active-target nucleus such as  $^{20}$Ne, $^{36,38,40}$Ar, or $^{80,82,83,84,86}$Kr,
or $^{28}$Si.

\subsection{Production of  a toroidal High-Spin Isomer by deep-Inelastic scattering}

\begin{figure} [t]
\includegraphics[angle=0,scale=0.55]{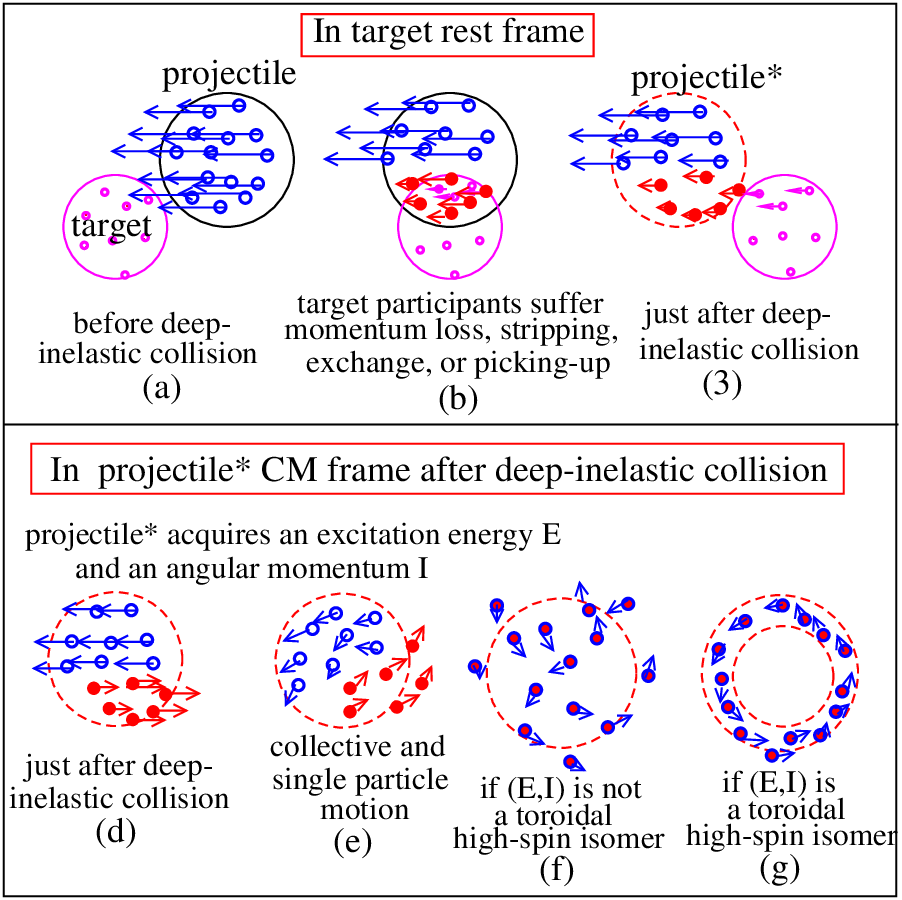} 
\vspace*{0.0cm}
\caption{(Color online) A schematic example of how a toroidal
  high-spin isomer of an excited projectile may be produced by a
  deep-inelastic heavy-ion scattering between a projectile nucleus and
  a target nucleus, showing a cut in the collision plane.  We display
  (a)-(c) in the target nucleus rest frame,
  and we show (d)-(g) in the excited
  projectile frame.  The time sequence proceeds from
  (a)-(e).  How (e) evolves
  to (f) or (g) depend on whether the
  excitation energy $E$ and spin $I$=$I_z$ are the same as those of
  the toroidal high-spin isomers.}
\label{deep}
\end{figure}

In a deep-inelastic collision between two heavy-ions at an energy near
the fermi energy, there are reaction products consisting of highly
excited binary systems with large angular momenta, in regions that may
be kinematically separable
\cite{Hag94,Lar96,Wue09,Wad04,Lin14,Cas87,Ono99,Cha10,Man10,Lac04}.
This led Cao $et~al.$}  \cite{Cao18} to suggest the use of such deep-inelastic collisions for
the production of light toroidal high-spin isomers.

We present below a schematic description how the deep-inelastic
collision mechanism may lead to a toroidal high-spin isomer, if
$(E,I)$ is appropriately the same as that of the toroidal high-spin
isomer.  We envisage that in a deep inelastic scattering in the target
rest frame in Fig.\ \ref{deep}, the semi-peripheral collision allows
the spectator nucleons of the projectile to stream forward, 
while the participant projectile nucleons collide with target nucleons 
and suffer a deceleration, as is depicted
in Fig.\ \ref{deep}(a)-(c).  As a consequence, the excited projectile
nucleus that emerges after the deep-inelastic collision is a spinning
nucleus in its own center-of mass frame.  It acquires both a
collective cranking motion and particle-hole excitations at an
excitation energy $E$ and an angular momentum $I$.  Its evolution can
be schematically depicted as in Fig.\ \ref{deep}(d)-(g).  When the
excitation energy $E$ and the angular momentum $I$ of the emerging
excited projectile are not those of a toroidal isomer, the excited
projectile nucleus will break up as described in statistical models
\cite{Ono99,Cas87,Cha10,Man10,Lac04}.  On the other hand, when the
excitation energy $E$ and the angular momentum $I$ of the emerging
projectile corresponds to that of a toroidal high-spin isomer, the
collective cranking motion and the rearrangement of the
single-particle motion of the nucleons may eventually settle down into
the toroidal shape of the high-spin isomer self-consistently because
of the conservation of energy and angular momentum, and the many-body
final-state interactions.

\subsection{Production of a toroidal vortex nucleus by punching through a target nucleus}

\begin{figure} [t]
\includegraphics[angle=0,scale=0.55]{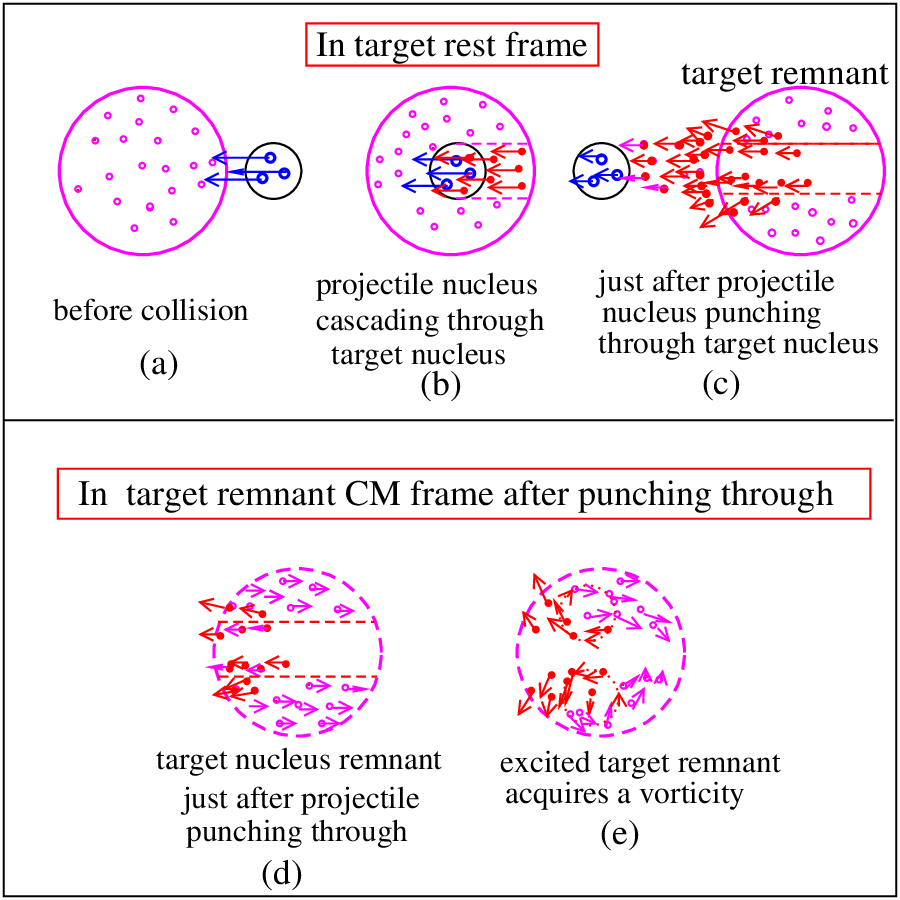} 
\vspace*{0.0cm}
\caption{(Color online) A schematic example of how a toroidal vortex
  isomer of the excited target remnant may be produced by the punching
  through of a smaller projectile on a larger target nucleus in an
  energetic nearly head-on collision, showing a cut in the collision
  plane.  We display  (a)-(c) in the
  target nucleus rest frame, while we show
  (d)-(e) in the excited remnant CM
  frame.  }
\label{vorpro}
\end{figure}

It has been suggested that a toroidal nucleus may be produced by
punching an energetic smaller heavy ion nearly head-on through a
larger target nucleus \cite{Won73}.  Scattering and evaporation takes
place within a small cone of the incident ion.  The ``remnant'' after
prompt cascade and evaporation may have a hole in the middle and
consequently have the geometry of a torus as depicted in
Fig.\ \ref{vorpro}.  After the incident projectile ion has punched
through the target nucleus, the target nucleons in the interaction
regions receive a momentum kick and have a velocity different from the
spectator target nucleons.  In the center-of-mass system of the target
remnant nucleus, vorticity will be developed as depicted in
Fig.\ \ref{vorpro}.  Indeed, vorticity has been found theoretically in
the hydrodynamical and model calculations in the collision of
high-energy heavy-ions \cite{Pan16,Den16,Iva17,Iva18}.  Vorticity in
heavy-ion collisions reveal two vortical structures that are common in
many fluid dynamic systems.  The vorticity and pairing of longitudinal
vortices with opposite signs are generated in the transverse plane.
The punching through of a large nucleus to form a toroidal nucleus
\cite{Won73} may therefore be a promising mechanism for the generating
isolated toroidal vortex nuclei.

In addition to using energetic small heavy-ion to punch through a
target nucleus one can also use the collision of an antineutron or
antiproton to punch through the target nucleus for the production of a
toroidal nucleus with a vorticity.  We envisage that at appropriate
collision energies, annihilation of the antiparticle takes place
inside the nucleus, and the momentum of the incident high-energy
antiparticle carries the produced particle forward in the form of a
cone, with the possibility of creating a hole inside the target
nucleus.  Among the remnants from such a collision, some target
nucleons receive a momentum kick from the produced particles and a
toroidal nucleus with vorticity may be created in a way similar to
what is depicted in Fig. \ref{vorpro}.

Toroidal galaxies (ring galaxies) have been known for some time
\cite{ringgal,wikiring,The76,Mad09,Rap10}.  It is generally held that
many of these ring galaxies arise from the collision of two galaxies,
and the catalogue of Ref.\ \cite{Mad09} list 127 observed  ring collision
galaxies.  The most well-known example is the
Arp147 which is composed of newly formed bright stars arising from the
collision of one smaller galaxy through  another larger galaxy.  When the two galaxies
collided, they pass though each other and the gravitational wave from
the impact leads to the condensation of the gas and dust into stars
\cite{Rap10}.  The observations of the ultraluminous X-ray sources in
the ring galaxy Arp147 in a ring of beads confirm the conventional
wisdom that collisions of gas-rich galaxies trigger large rates of
star formation which, in turn, generate substantial numbers of X-ray
sources, some of which have luminosities above the Eddington limit for
accreting stellar-mass black holes as ultraluminous X-ray sources
\cite{Rap10}.  Another example is Arp148 showing a ring with a jet of
a smaller galaxy along the symmetry axis of the ring \cite{wikiring}.
The existence of toroidal galaxies is only suggestive that if toroidal
vortex nucleus could be formed in the collision of a light nucleus
with a heavy nucleus in a nearly head-on collision, then it would
exist.

\section{Conclusions and Discussions}

We study here the nature of the toroidal single-particle states and
their wave functions in the intermediate mass region where negative
shell corrections are expected.  We find that toroidal shells occur with a regular structure 
 in
an extended region with many toroidal magic numbers at various
toroidal deformations $R/d$.  The enhanced stability associated with the
nuclear shell effects suggests that there may be many exited toroidal
states stable against expansion and contraction of the toroidal major
radius.

Toroidal Routhian energies under the constraint of an angular momentum
have been evaluated, and one finds regions of low Routhian energy
density indicating that toroidal high-spin isomers may also have a
common occurrence in the intermediate mass region.

A new vorticity degree of freedom opens up for examination for
toroidal nuclei in the intermediate mass region.  There can be vortex
creating particle-hole excitations that will allow the nucleus to become a toroidal vortex nucleus with 
a net vorticity by promoting nucleons from states of vorticity of one
sign to selectively populate un-occupied states with vorticities of
the other sign.

 To get more accurate locations of the toroidal shells, a more general
 potential such as those with a Wood-Saxon shape potential will be
 useful in determining the toroidal shell nucleon numbers and
 high-spin isomers.  Future mean-field calculations in the
 intermediate mass region will be of great interest to give reliable
 quantitative estimates of the energies of the toroidal isomers.
 From all indications and results from the present work, the
 intermediate mass region hold the promise to be a rich region for the
 exploration of the toroidal degree of freedom.

There remain many interesting problems that will need to be considered
in the future.  The question of the decay of the toroidal nucleus, the
stability against sausage distortions or the Plateau-Rayleigh
instability \cite{plateau,rayleigh14,eggers97}, and the effects of the
quantization of the spin on the sausage instability, will need to be
addressed.  There are also the questions of pairing interaction that
may be weakened by the cranking of a pair having opposite tendencies
in changing the energy, and the questions on the effects of
self-consistent mean-fields and the signatures for toroidal nuclei.
These and many other unresolved questions make the study of toroidal
nucleus a very interesting area for further investigation.

\vspace*{0.3cm}
\centerline{\bf Acknowledgments}
\vspace*{0.3cm}

  The authors would like to thank Prof. Yitzhak Sharon for helpful discussions. The research was supported in part by the Division of Nuclear
  Physics, U.S. Department of Energy under Contract DE-AC05-00OR22725.

\appendix

\section{Solution of the single-particle Hamiltonian}

We would like to show how the eigenenergy of the single-particle state
in Eq.\ (\ref{eq6}) can be approximately solved.  We
split the spin-orbit interaction of Eq.\ (\ref{eq5}) into two parts:
\begin{eqnarray}
&&- \frac{2\kappa \hbar}{m \oo \omega} \bb s \cdot \left ( \nabla V_0 \times \bb p \right ) =V^{\rm dia}_{\rm so}+V^{\rm off}_{\rm so},
\nonumber\\
V^{\rm dia}_{\rm so}&&=- \frac{\kappa \hbar}{m \oo \omega} 
 \begin{pmatrix}
 1 & 0 \cr 0 & -1 
 \end{pmatrix} 
   \frac{\partial V_0}{\partial \rho} \frac{\hbar \partial}{i\rho \partial \phi} ,
\nonumber\\
V^{\rm off}_{\rm so}&&= - \frac{\kappa \hbar}{m \oo \omega }
\biggl  \{ 
   \begin{pmatrix}
  0 & e^{-i\phi} \cr 
   e^{i\phi}  & 0
 \end{pmatrix}
\left (  - \frac{\partial V_0}{\partial z} \frac{\hbar \partial}{i\rho \partial \phi} \right )
\nonumber\\
&&
+
 \begin{pmatrix}
  0 & -e^{-i\phi} \cr
    e^{i\phi} & 0
 \end{pmatrix}
   \biggl ( -\frac{\partial V_0}{\partial \rho}\frac{\hbar \partial}{ \partial z} 
+  \frac{\partial V_0}{\partial z}\frac{\hbar\partial}{ \partial \rho} \biggr )
\biggr  \}.
\end{eqnarray}
For simplicity, we shall neglect the small contribution from the
off-diagonal spin-orbit interaction as it gives higher-order
transition matrix elements with a relatively large energy denominator.
We consider therefore the approximate Hamiltonian
\begin{eqnarray}
H_0&&=-\frac{\hbar ^2}{2m} \nabla^2 + V_0(\rho,z)+V_{\rm so}^{\rm dia},
\end{eqnarray}
where 
\begin{eqnarray}
V_{\rm so}^{\rm dia}= - \frac{2\kappa \hbar}{m \oo \omega} (\frac{1}{\rho}\frac{\partial V_0}{\partial \rho})
s_z L_z&&= - \frac{2\kappa \hbar \omega_\perp^2}{ \oo \omega}  \frac{\rho-R}{\rho}s_z \frac{\hbar \partial}{i \partial \phi}.~~~~~
\end{eqnarray}
We solve for eigenstates $|n_\rho n_z \Lambda_z  \Omega_z\rangle$ of $H_0$,
\begin{eqnarray}
H_0 | n_\rho n_z \Lambda_z \Omega_z \rangle
= \epsilon(n_\rho n_z \Lambda _z\Omega _z) | n_\rho n_z \Lambda_z \Omega_z  \rangle,
\end{eqnarray}
normalized according to 
\begin{eqnarray}
\int dz ~\rho d\rho ~d \phi | {\hat {\cal R} }_{n_\rho}(\rho) Z_{n_z}(z) \Phi_{\Lambda_z}(\phi) |^2\chi_{s_z}^{\dagger} \chi_{s_z}=1.
\end{eqnarray}
The two states with $\Omega_z=\Lambda_z+s_z=\pm  |\Omega_z|$ have the same energy.  They are degenerate.
Writing out the operator explicitly, we get 
 \begin{eqnarray}
\biggl [  -\frac{\hbar ^2}{2m} ( \frac{1}{\rho} \frac{\partial}{\partial  \rho} \rho  \frac{\partial}{\partial  \rho}
- \frac{\Lambda_z^2}{\rho^2}) 
&+\frac{1}{2} m \omega_\perp^2 (\rho -R)^2+V_{\rm so}^{\rm dia}
  \biggr ]{\hat{ \cal R}}_{n_\rho}(\rho)
 \nonumber\\
 &= \epsilon_{n_\rho \Lambda_{n_z} \Omega_z}^0{\hat{ \cal R}}_{n_\rho}(\rho),
\label{eq13}
\end{eqnarray}
 \begin{eqnarray}
\biggl [  -\frac{\hbar ^2}{2m}  \frac{\partial^2}{\partial  z^2}
+\frac{1}{2} m \omega_\perp^2 z^2
 - \epsilon_{n_z}^0 \biggr ] Z_{n_z}(z)=0,
\label{eqz}
\end{eqnarray}
 \begin{eqnarray}
\biggl [  i  \frac{\partial}{\partial  \phi}
 - \Lambda_z \biggr ] \Phi_{\Lambda_z}(\phi)=0.
\end{eqnarray}
Single-particle eigenvalue is
\begin{eqnarray}
 \epsilon(n_\rho n_z \Lambda_z \Omega_z) = \epsilon_{n_\rho \Lambda_z \Omega_z}^0+\epsilon_{n_z}^0.
\label{a9}
\end{eqnarray}
We can transform ${\hat{ \cal R}}_{n_\rho}(\rho)$ in Eq.\ (\ref{eq13}) by
\begin{eqnarray}
 {\cal R}_{n_\rho}(\rho) = \sqrt{\rho}{\hat  {\cal R}}_{n_\rho}(\rho).
\end{eqnarray}
We normalize the single-particle wave function according to 
\begin{eqnarray}
\int dz \!|  Z_{n_z}(z) |^2 &=&\!\!\int \!\! d\rho  | {\cal R}_{n_\rho}(\rho) |^2
=\int \!\!  d \phi | \Phi_{\Lambda_z}(\phi) |^2 
\nonumber\\
&=& \chi_{s_z}^{\dagger} \chi_{s_z}
=1.
\nonumber
\end{eqnarray}
We obtain from Eq.\ (\ref{eq13})
\begin{eqnarray}
 && \biggl [   -\frac{\hbar ^2}{2m}
{\frac{\partial^2}{\partial \rho^2}}
+ \frac{\hbar ^2}{2m} \frac{\Lambda_z^2-\frac{1}{4}}{\rho^2}
+\frac{1}{2} m \omega^2 (\rho -R)^2 +V_{\rm so}^{\rm dia} 
  \biggr ]{\cal R}(\rho)
\nonumber\\
 &&~~~~~~~~~~~~~~~~~~~~~~~~~~~~~~~~= \epsilon_{n_\rho \Lambda_z \Omega_z}^0 {\cal R}(\rho)
\label{eq17}
\end{eqnarray}
It is useful to make a change of coordinates 
\begin{eqnarray}
q= \rho-R.
\end{eqnarray}
We expand $\rho$ about $R$ in power of $q$ and keep terms up to the
second order in $q/R$.  Eq.\  (\ref{eq17}) becomes
\begin{eqnarray}
 &&\biggr \{\!\!\!   - \! \frac{\hbar ^2}{2m}
{\frac{\partial^2}{\partial q ^2}}
\!+\! \frac{\hbar ^2}{2m} \frac{\Lambda^2 \!-\!\frac{1}{4}}{R^2}\!
\left [\! 1 \!\!-\!\! \frac{2q}{R}
\!+\!\frac{3q ^2}{ R^2}\right ]
\!\!+\!\!\frac{m \omega^2 q^2}{2}\!+\!V_{\rm so}^{\rm dia} 
\!\!  \biggr \}\!{\cal R}(\rho)
\nonumber\\
 &&~~~~~~~~~~~~~~~~~~~~~~~~~~~~~~~~= \epsilon_{n_\rho \Lambda_z \Omega_z}^0 {\cal R}(\rho)
\label{eqr}
\end{eqnarray}
We can solve the approximate Hamiltonian in $q$.  We have
\begin{eqnarray}
s_z L_z 
= \frac{J_z^2 -\Lambda_z^2 - 1/4}{2}= \frac{\Omega_z^2 -\Lambda_z^2 - 1/4}{2}.
\end{eqnarray}
We assume  the large radius $R$ approximation with small $d/R$, and expand 
 $(\frac{1}{\rho}\frac{\partial V_0}{\partial \rho})$
 in powers of $q$ up to $q^2$,
\begin{eqnarray}
 (\frac{1}{\rho}\frac{\partial V_0}{\partial \rho})=m \omega_\perp^2 \frac{(\rho -R)}{\rho}
 =  m \omega_\perp^2\left (\frac{ q}{R}-\frac{q^2}{R^2}\right ) .
\end{eqnarray}
We also expand $(\Lambda^2-1/4)/\rho^2$ in power of $q$ up to the second order of $q$,
\begin{eqnarray}
 \frac{\Lambda_z^2-\frac{1}{4}}{\rho^2}=\frac{\Lambda_z^2-\frac{1}{4}}{R^2}
\left [ 1 -\frac{2q}{R}
+\frac{3q ^2}{ R^2}\right ].
\end{eqnarray}
The eigenvalue equation can be rewritten  as 
\begin{eqnarray}
 &&\biggl \{   -\frac{\hbar ^2}{2m}
{\frac{\partial^2}{\partial q ^2}}
+ \frac{\hbar ^2}{2m} \frac{\Lambda_z^2-\frac{1}{4}}{R^2}
+\frac{1}{2} m \omega_\perp^2 q^2 
+V^{(1)}
\nonumber\\
&&~~~
 - \epsilon_{n_\rho \Lambda_z \Omega_z}^0 \biggr \}{\cal R}(\rho)
 =0
 \label{46}
\end{eqnarray}
where
\begin{eqnarray}
V^{(1)}(q)&&=+ \frac{\hbar ^2}{2m} \frac{\Lambda_z^2-\frac{1}{4}}{R^2}
\left (  -\frac{2q}{R}
+\frac{3q ^2}{ R^2}\right )
\nonumber\\
&&
 -  \frac{2\kappa (\hbar \omega_\perp)^2}{\hbar \oo  \omega} 
s_z L_z
\left ( \frac{q}{R} - \frac{q^2}{R^2} \right ).
\nonumber
\end{eqnarray}
We cast $  \frac{1}{2} m \omega^2 q^2$+$V^{(1)}(q)$ into a displaced quadratic form
\begin{eqnarray}
\frac{1}{2} m \omega^2 q^2+V^{(1)}(q)=\frac{1}{2}m\omega_\perp '^2 (q-q_0)^2+ a_0 ,
\end{eqnarray}
where 
\begin{eqnarray}
\omega_\perp'^2&&=\omega_\perp^2(1+a_2), 
\end{eqnarray}
$a_0$, $a_2$, and $q_0$ are given by 

\begin{eqnarray}
a_2&&=\frac{\hbar ^2}{2m} \frac{\Lambda_z^2-\frac{1}{4}}{R^2}
\frac{3}{ \frac{1}{2}m\omega_\perp^2 R^2}
 +\frac{2\kappa (\hbar \omega_\perp)^2}{\hbar \oo  \omega} 
  \frac{s_z \Lambda_z}{\frac{1}{2}m\omega_\perp^2 R^2},~~~~~
\\
a_0&&=-\frac{1}{2} m\omega_\perp'^2 q_0^2 ,
\\
q_0&&=-\frac{1}{2}a_1R\left (\frac{\omega_\perp}{\omega_\perp'}\right )^2,
\\
a_1&& =-\frac{\hbar ^2}{2m} \frac{\Lambda_z^2-\frac{1}{4}}{R^2}  \frac{2}{ \frac{1}{2}m\omega_\perp^2 R^2}
 \frac{s_z \Lambda_z}{\frac{1}{2}m\omega_\perp^2 R^2}.
\end{eqnarray}

Eq. (\ref{46}) becomes
\begin{eqnarray}
&& \biggl \{   -\frac{\hbar ^2}{2m}
{\frac{\partial^2}{\partial q ^2}}
+ \frac{\hbar ^2}{2m} \frac{\Lambda_z^2-\frac{1}{4}}{R^2}
+\frac{1}{2} m \omega_\perp'^2 (q-q_0)^2
+a_0~~~~~
\nonumber\\
&&~~~~~~~~~~~~~~~~~~~~~~~~~~~~~~~
 - \epsilon_{n_\rho \Lambda_z \Omega_z}^0 \biggr \}{\cal R}(\rho)=0.
\end{eqnarray}
The above eigenenergy solution for the $\rho$ degree of freedom is
\begin{eqnarray}
\epsilon^0_{n_\rho \Lambda_z \Omega_z}= \frac{\hbar ^2}{2m} \frac{\Lambda_z^2-\frac{1}{4}}{R^2} +(n_\rho+\frac{1}{2})\hbar \omega_\perp' +a_0.
\end{eqnarray}
After substituting the above eigenvalue into Eq.\ (\ref{a9}), we get the single particle energy given by Eq.\ (\ref{eq6}),
\begin{eqnarray}
 \epsilon(n_\rho n_z \Lambda_z \Omega_z)&& =(n_\rho+\frac{1}{2})\hbar \omega_\perp' + (n_z+\frac{1}{2})\hbar \omega_\perp
  \nonumber\\
&&+  \frac{\hbar ^2}{2m} \frac{\Lambda_z^2-\frac{1}{4}}{R^2} +a_0.
\label{eq19}
\end{eqnarray}

\end{document}